\documentclass{PoS}
\usepackage{bm}% bold math
\usepackage{amsmath}% bold math
\usepackage{colortbl}
\usepackage{epstopdf}    % converts eps fig to pdf format
\usepackage{wrapfig}     % wraps fig in text
\usepackage{caption}
\usepackage{subcaption}
\usepackage{multirow}
\usepackage{amsmath}
\usepackage{cleveref}
\usepackage{array,hhline}
\title{Dynamical net charge fluctuations at RHIC\\
 energies in STAR}

\ShortTitle{Dynamical net charge fluctuations in STAR}

\author{\speaker{Bhanu SHARMA (for the STAR Collaboration)}\\
         Panjab University, Chandigarh, India\\
        E-mail: \email{bhanu.pu@gmail.com}}

\abstract{We present preliminary results from the study of dynamical net-charge fluctuations measured in Au+Au collisions at $\sqrt{s_{NN}}$ = 39 GeV to 7.7 GeV and compare with the published results at higher collision energies. We also present the centrality dependence of the dynamical net-charge fluctuations.}

\FullConference{7th International Conference on Physics and Astrophysics of Quark Gluon Plasma\\
		1-5 February , 2015\\
		Kolkata, India}

\begin{document}

\section{Introduction}
The STAR experiment at the Relativistic Heavy Ion Collider (RHIC) provides the ability to investigate the behaviour of strongly interacting matter at high density and to study the Quark Gluon Plasma (QGP). In the year 2010 (Run 10), RHIC started its Beam Energy Scan (BES) program and collided Au+Au ions from $\sqrt{s_{NN}}$ = 39 GeV down to 7.7 GeV covering 112 $<$ $\mu_{B}$ $<$ 410 MeV. This allows one to access and probe broad regions of the QCD phase diagram. Event-by-event net-charge fluctuations have been proposed as an indicator of the QGP formation in heavy ion collisions. The fluctuation in net-charge depends on the squares of the charges present in the system, which depends on the state from which it originates. The system passing through a QGP phase which has quarks as charge carriers should result in a significantly different net-charge fluctuation as compared to Hadron Gas (HG). The variance of the event-by-event difference of the numbers of positive and negative particles scaled by the total charged particle multiplicity, a quantity called $\textit{D}$, should be approximately four times smaller in a QGP medium than in a gas of hadrons. The charge  fluctuation is measured in terms of $ \textit{D} $ defined as :
\begin{equation}
D=4\dfrac{\langle\delta Q^{2}\rangle}{\langle N_{ch}\rangle},
\end{equation}
where $\langle\delta Q^{2}\rangle$ is the net charge variance, $Q=N_{+}-N_{-}$ and $N_{ch}=N_{+}+N_{-}$. Here $N_{+}$ and $N_{-}$ are the number of negative and positive particles, measured in specific transverse momentum $(p_{T})$ and pseudorapidity $(\eta)$ window.\\
\indent The value of $\textit{D}$ is 4 for an uncorrelated pion gas and is reduced by about 30$\%$ in the presence of resonances. For a thermal system of free quarks and gluons, the value  is significantly lower and has been calculated to be $\approx$ 1. \cite{ebye_1,ebye_2}\\
\indent The event-by-event net-charge fluctuations have also been estimated  by calculating the quantity $\nu_{+-,dyn}$  defined as: \cite{b2,pruneau,monika,prc} :
%%%%%%%%%%%%%%%%%%%%%%%%%%%%%%%%%% equation %%%%%%%%%%%%%%%%%%%%%%%%%%%%%%%%%%%%
\begin{equation}
\nu_{+-,dyn}=\dfrac{\langle N_{+}(N_{+}-1)\rangle}{\langle N_{+}\rangle^{2}}+\dfrac{\langle N_{-}(N_{-}-1)\rangle}{\langle N_{-}\rangle^{2}}
-2\dfrac{\langle N_{-}N_{+}\rangle}{\langle N_{+}\rangle\langle N_{-}\rangle}
\end{equation}
which is a measure of the relative correlation of $+$$+$, $-$ $-$ and $+$$-$ pairs. The $\nu_{+-,dyn}$ has been found to be robust against experimental inefficiencies \cite{robust}. The value of \textit{D} is related to $\nu_{+-,dyn}$ as :
\begin{equation}
\langle N_{ch}\rangle \nu_{+-,dyn}=D-4 .
\end{equation}
\section{Analysis Details and Results}
The measurements of net-charge fluctuations as a function of centrality in Au+Au collisions at $\sqrt{s_{NN}}$ = 7.7, 11.5, 19.6, 27 and 39 GeV data collected in years 2010 and 2011 by the STAR experiment are reported. For this analysis, we use charged 
\begin{figure}
\begin{center}
\includegraphics[width=80mm]{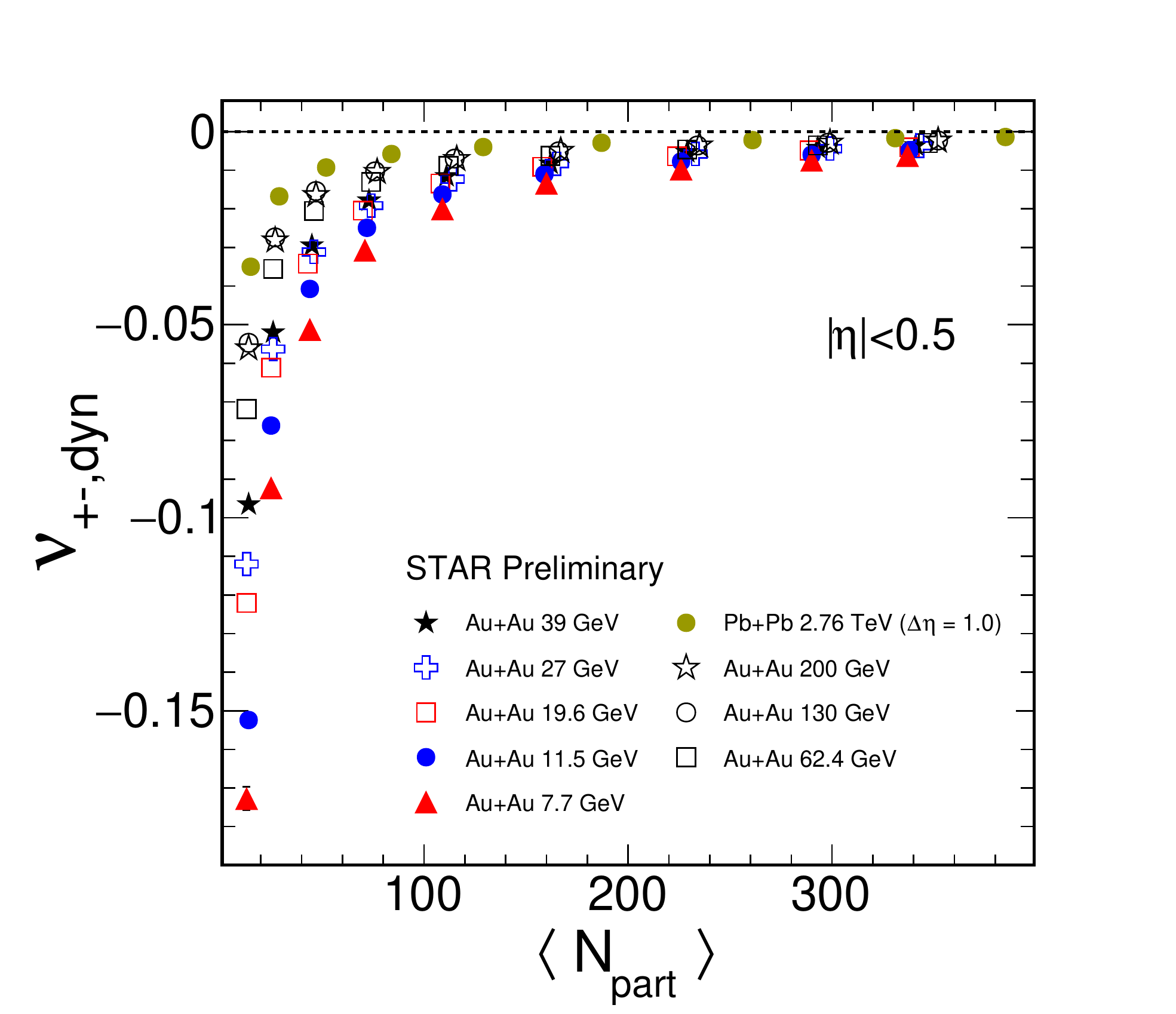}
\caption{ Dynamical net-charge fluctuations, $\nu_{+-,dyn}$, of charged particles as a function of number of participating nucleons, $\langle N_{part}\rangle$.}
\label{fig1}
\end{center}
\end{figure}
particle tracks from the Time Projection Chamber (TPC) for -0.5 $<$ $\eta$ $<$ 0.5  with  transverse momenta in the range 0.2 $<$ $p_{T}$ $<$ 5.0 GeV/$c$. For the 
\begin{figure}
\begin{center}
\includegraphics[width=80mm]{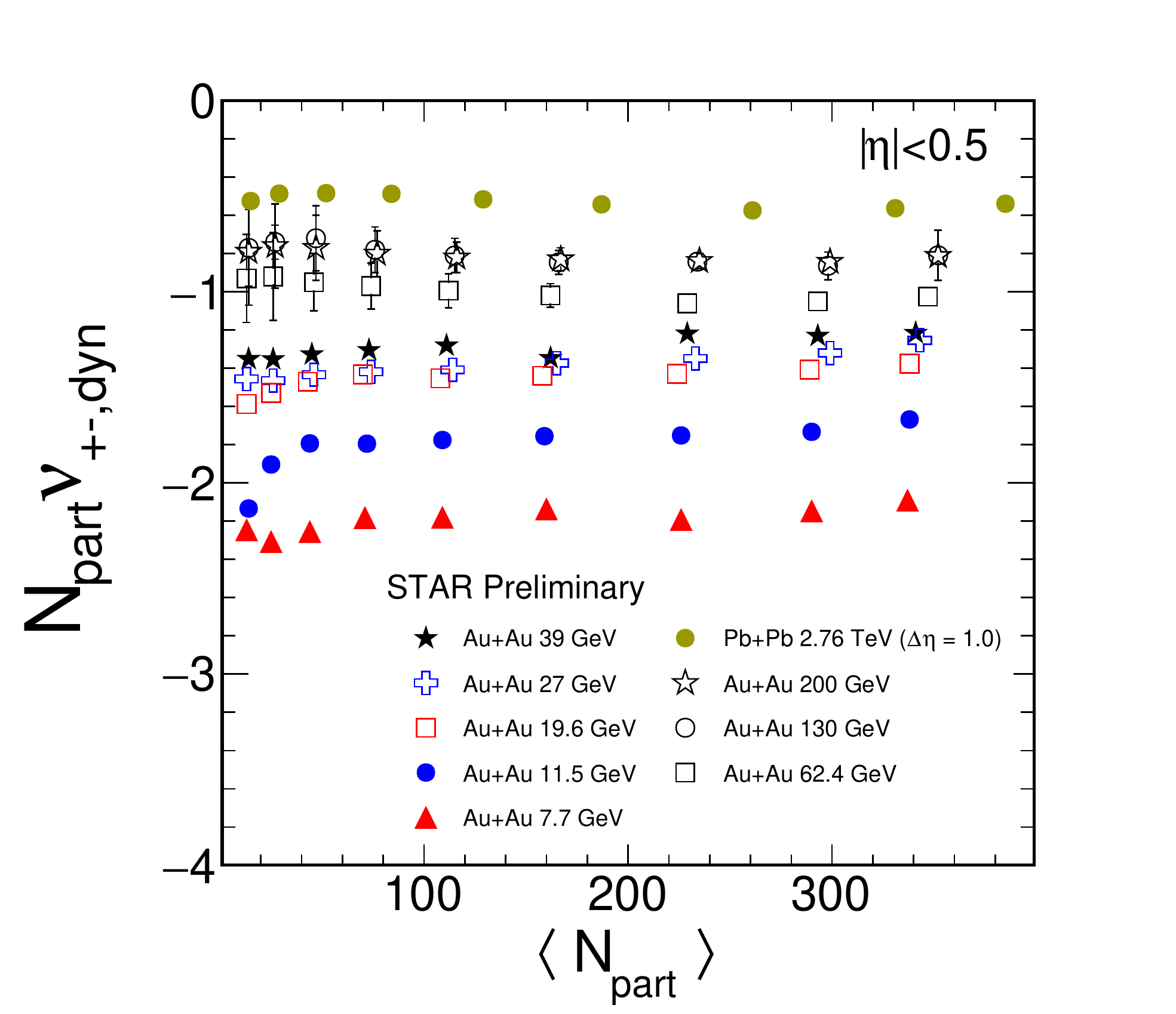}
 \caption{Dynamical net-charge fluctuations, $\nu_{+-,dyn}$, scaled with the average number of participating nucleons as a function of $\langle N_{part}\rangle$.}
\label{fig2}
\end{center}
\end{figure}
centrality selection, the uncorrected multiplicity of charged particles within 0.5 $<$ $|\eta|$ $<$ 1.0 is used in order to avoid the autocorrelation between the centrality definition and the $\nu_{+-,dyn}$ observable.\\
\indent Figure~\ref{fig1} shows the measurements of $\nu_{+-,dyn}$ at $\sqrt{s_{NN}}$ = 7.7, 11.5, 19.6, 27 and 39 GeV as a function of the average number of participating nucleons, $\langle N_{part}\rangle$. It also includes the previous results from STAR at $\sqrt{s_{NN}}$ = 200, 130 and 62.4 GeV, and ALICE results for Pb+Pb collisions at $\sqrt{s_{NN}}$ = 2.76 TeV \cite{monika,alice}. The uncertainties shown are only statistical. In all cases, the values of $\nu_{+-,dyn}$ are negative indicating the dominance of the correlation between positive and negative charged particles. The strength of the correlation decreases while going from peripheral to central collisions. Also, the magnitude of fluctuations decreases as the beam energy increases.\\ 
\indent Figure~\ref{fig2} shows $\nu_{+-,dyn}$ scaled with the average number of participating nucleons, $N_{part}\nu_{+-,dyn}$, as a function of $\langle N_{part}\rangle$. The measured data scaled by the number of participating nucleons exhibits either a weak or no centrality dependence at all of the measured energies.\\
\indent The magnitude of the net-charge correlations is affected by the global charge 
\begin{figure}
\begin{center}
\includegraphics[width=100mm]{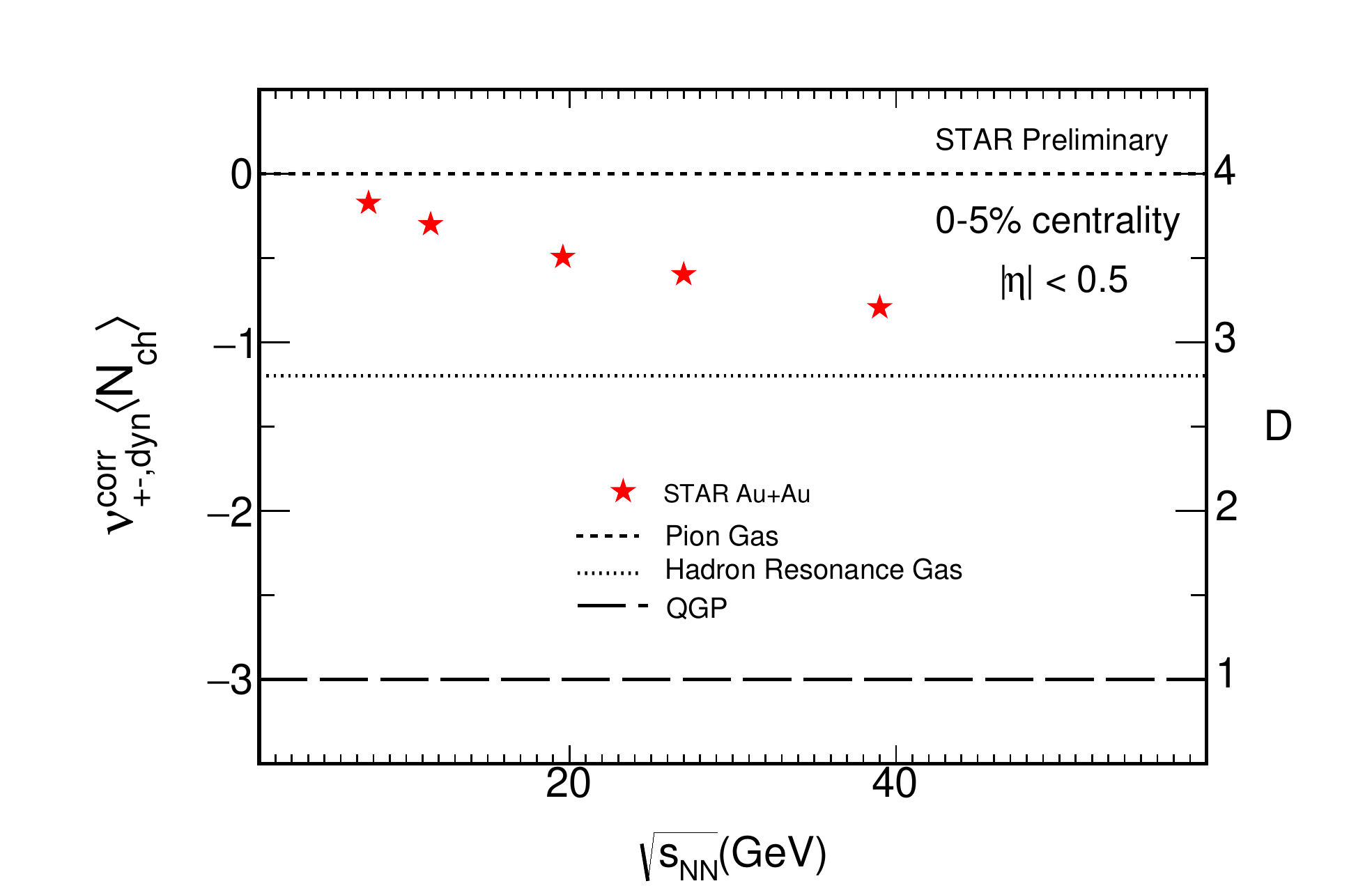}
\caption{$\langle N_{ch}\rangle \nu_{+-,dyn}^{corr}$ (left-axis) and $D$ (right-axis) as a function of beam energy for 0-5\% collisions. The theoretical predictions for a Pion Gas, HG and a QGP are also indicated. Systematic uncertainties are not yet included.}
%\caption{\label{fig2} $\langle N_{ch} \rangle × \nu_{+−,dyn}$ (left-axis) and D (right-axis) plotted for $|\eta|$ $<$ 0.5 as a function of the number of participating nucleons.}
\label{fig3}
\end{center}
\end{figure}
conservation and the finite size of the colliding system. The contribution of the global charge conservation effect is  estimated to be -4/$\langle N_{total}\rangle$, \cite{pruneau} where $\langle N_{total}\rangle$ is the average number of charged particles produced over the full phase space. The corrected value of $\nu_{+-,dyn}$ is calculated as
\begin{equation}
\nu_{+-,dyn}^{corr} = \nu_{+-,dyn} + \dfrac{4}{\langle N_\mathrm{{total}}\rangle}
\end{equation}
%The $\nu_{+-,dyn}^{corr}$ is related to $D$ as :
%\begin{equation}
%D = 4 + \langle N_{ch}\rangle \nu_{+-,dyn}^{corr}
%\end{equation}
Here the $\langle N_{ch}\rangle$ is the efficiency corrected average charged  particle multiplicity for Au+Au collisions. Both $\langle N_{ch}\rangle$ and $\langle N_{total}\rangle$ have been estimated from the PHOBOS experiment data \cite{phobos_ntot}. Figure~\ref{fig3} shows $\langle N_{ch}\rangle \nu_{+-,dyn}^{corr}$ along left y-axis and $D$ along right y-axis as a function of the colliding energy for 0-5\% central collisions within $|\eta|$ $<$ 0.5. The uncertainties shown are statistical only. The theoretical predictions for Pion Gas, Hadron Resonance Gas and QGP have also been displayed in the figure. It can be observed that the net-charge fluctuations corrected for the global charge conservation when scaled with $\langle N_{ch}\rangle$ decrease with increases in the beam energy.  The $D$ measure linearly decreases with increasing energy. The net-charge fluctuations observables approach the expectation for a Pion Gas as the beam energy decreases.\\
\indent In summary, we report recent results of the net-charge fluctuations for Au+Au collisions at 7.7, 11.5, 19.6, 27 and 39 GeV. There is a monotonic reduction in the magnitude of dynamical net-charge fluctuations with increasing number of participants. Dynamical net-charge fluctuations are observed to follow approximate $N_{part}$ scaling. Top 5\% central collisions results show that $\langle N_{ch}\rangle \nu_{+-,dyn}^{corr}$ generally decreases with increasing colliding energy.

\end{document}